\documentclass[twocolumn,prb,showpacs]{revtex4}
\usepackage{bm}
\usepackage{epsf}

\begin{document}
\title{Spin-dependent tunnelling through a symmetric barrier}
\author{V.I.~Perel'}
\author{S.A.~Tarasenko}\email{tarasenko@coherent.ioffe.rssi.ru}
\author{I.N.~Yassievich}
\affiliation{A.F.~Ioffe Physico-Technical Institute, 194021
St.Petersburg, Russia}
\author{S.D.~Ganichev}
\author{V.V.~Bel'kov}
\author{W.~Prettl}
\affiliation{Fakult\"{a}t f\"{u}r Physik, Universit\"{a}t
Regensburg,  D-93040 Regensburg, Germany}
\begin{abstract}
The problem of electron tunnelling through a symmetric
semiconductor barrier based on zinc-blende-structure material is
studied. The $k^3$ Dresselhaus terms in the effective Hamiltonian
of bulk semiconductor of the barrier are shown to result in a
dependence of the tunnelling transmission on the spin orientation.
The difference of the transmission probabilities for opposite spin
orientations can achieve several percents for the reasonable width
of the barriers.
\end{abstract}
 \maketitle

Lately spin polarized electron transport in semiconductors
attracts a great attention.~\cite{spintronic} One of the major
problems of general interest is a possibility and methods of spin
injection into semiconductors. A natural way to achieve spin
orientation in experiment is the injection of spin polarized
carriers from magnetic materials. Although significant progress
has been made recently,~\cite{Zhu,Hammar,Hanbicki,Safarov}
reliable spin-injection into low-dimensional electrons systems is
still a challenge. Schmidt et al. pointed out that a fundamental
obstacle for electrical injection from ferromagnetic into
semiconductor was the conductivity mismatch of the metal and the
semiconductor structure.~\cite{Schmidt} However, Rashba showed
that this problem could be resolved by using tunnelling contact at
the metal-semiconductor interface.~\cite{Rashba2} On the other
hand Voskoboynikov et al.~\cite{Vosk1} proposed that asymmetric
non-magnetic semiconductor barrier itself could serve as a spin
filter. It was demonstrated that spin-dependent electron
reflection by inequivalent interfaces resulted in the dependence
of the tunnelling transmission probability on the orientation of
electron spin. This effect is caused by interface-induced Rashba
spin-orbit coupling~\cite{Rashba} and can be substantial for
resonant tunnelling through asymmetric
double-barrier~\cite{Vosk2,Silva} or triple-barrier~\cite{Koga}
heterostructures. However, in the case of symmetric potential
barriers, the interface spin-orbit coupling does not lead to a
dependence of tunnelling on the spin orientation.

In this communication we will show that the process of tunnelling
is spin dependent itself. We demonstrate that a considerable spin
polarization can be expected at tunnelling of electrons even
through a single symmetric barrier if only the barrier material
lacks a center of inversion like zinc-blende structure
semiconductors. The microscopic origin of the effect is the
Dresselhaus $k^3$ terms~\cite{Dresselhaus} in the effective
Hamiltonian of the bulk semiconductor of the barrier.

\begin{figure}[t]
\epsfxsize=3in \epsfysize=1.6in \centering{\epsfbox{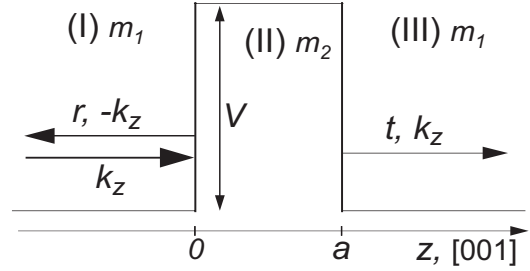}}
\caption{Electron tunnelling through the potential barrier $V$ of
width $a$.} \label{fig1}
\end{figure}

\begin{figure}[t]
\epsfxsize=3.8in \epsfysize=1.5in \centering{\epsfbox{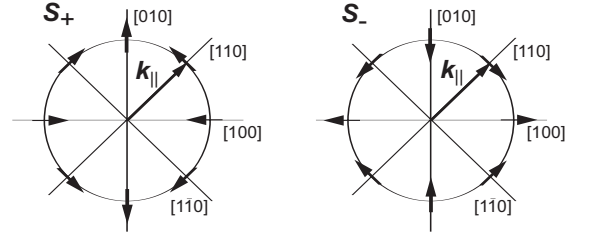}}
\caption{Spin orientation of eigen states "$+$" and "$-$" as a
function of the orientation of the in-plane electron wave vector
$\bm{k}_{\parallel}$.} \label{fig2}
\end{figure}

We consider the transmission of electrons with the initial wave
vector $\bm{k}=(\bm{k}_{\parallel},k_z)$ through a flat potential
barrier of height $V$ grown along $z \parallel [001]$ direction
(see Fig.~\ref{fig1}); $\bm{k}_{\parallel}$ is the wave vector in
the plane of the barrier, and $k_z$ is the wave vector normal to
the barrier pointing in the direction of tunnelling. The electron
Hamiltonian of the barrier in effective mass approximation
contains the spin-dependent $k^3$ term (Dresselhaus
term)~\cite{Dresselhaus}
\begin{equation}\label{H_D}
\hat{H}_D = \gamma \left[ \hat{\sigma}_x k_x (k_y^2-k_z^2) + \hat{\sigma}_y k_y (k_z^2-k_x^2) +
\hat{\sigma}_z k_z (k_x^2-k_y^2) \right] \:,
\end{equation}
where $\hat{\sigma}_{\alpha}$ are the Pauli matrices, $\gamma$ is
a material constant (see Table~1), and the coordinate axis $x,y,z$
are assumed to be parallel to the cubic crystallographic axis
$[100]$, $[010]$, $[001]$, respectively. In the case of tunnelling
along $z$ one should consider $k_z$ in the Hamiltonian as an
operator $-i \partial / \partial z$. We assume the kinetic energy
of electrons to be substantial smaller than the barrier high $V$,
then the Hamiltonian~(\ref{H_D}) is simplified to
\begin{equation}\label{H_D1}
\hat{H}_D = \gamma (\hat{\sigma}_x k_x - \hat{\sigma}_y k_y)
\frac{\partial^2}{\partial z^2}  \:.
\end{equation}
One can note that es\-sen\-tial\-ly $\hat{H}_D$ induces a
spin-\-de\-pen\-dent correction to the effective electron mass
along $z$ in the
barrier. \\
\begin{center}
\begin{tabular}[b]{|c|c|c|c|c|c|}
\hline  & GaSb & InAs & GaAs & InP & InSb \\ \hline $\gamma$,
eV$\cdot${\AA}$^{3}$ & 187 & 130 & 24 & 8 & 220 \\ \hline
$m^*/m_0$ & 0.041 & 0.023 & 0.067 & 0.081 & 0.013\\ \hline
\end{tabular}
\end{center}
\begin{center}
Table~1: Parameters of band structure of various A$_3$B$_5$
semiconductors.~\cite{PT,IP}
\end{center}
The Hamiltonian~(\ref{H_D1}) is diagonalized by spinors
\begin{equation}\label{chi_pm}
\chi_{\pm} = \frac{1}{\sqrt{2}}\left(
\begin{array}{c}
1 \\ \mp \mbox{e}^{- i \varphi}
\end{array}
\right) \:,
\end{equation}
which correspond to the electron states "$+$" and "$-$" of the
opposite spin directions. Here $\varphi$ is the polar angle of the
wave vector $\bm{k}$ in the plane $xy$, being
\begin{equation}
\bm{k} = (k_{\parallel} \cos \varphi \,, \: k_{\parallel} \sin
\varphi \,, \: k_z)  \:.
\end{equation}
Transmission probabilities for the electrons of eigen spin states
"$+$" and "$-$"~(\ref{chi_pm}) are different due to spin-orbit
term~(\ref{H_D1}). The orientations of spins $\bm{s}_\pm$ in the
states "$+$" and "$-$" depend on the in-plane wave vector of
electrons and are given by
\begin{equation}\label{spm}
\bm{s}_{\pm} = (\mp \cos \varphi \,, \: \pm \sin \varphi \,, \: 0)
\:.
\end{equation}
Fig.~\ref{fig2} demonstrates the orientations of spins
$\bm{s}_{+}$ and $\bm{s}_{-}$ for various directions of the
in-plane electron wave vector $\bm{k}_{\parallel}$, i.e. as a
function of polar angle $\varphi$. If $\bm{k}_{\parallel}$ is
directed along a cubic crystal axis ($[100]$ or $[010]$) then the
spins are parallel (or antiparallel) to $\bm{k}_{\parallel}$,
while $\bm{s}_{\pm}$ are perpendicular to $\bm{k}_{\parallel}$ if
the in-plane wave vector is directed along the axis $[1 \bar{1}0]$
or $[1 10]$.

Electrons of the eigen spin states "$+$" and "$-$" propagate
through the barrier with conserving of the spin orientation. Since
the wave vector in the plane of the barrier $\bm{k}_{\parallel}$
is fixed, wave functions of the electrons can be written in the
form
\begin{equation}
\Psi_{\pm}(\bm{r})=\chi_{\pm} u_{\pm}(z) \exp{(i
\bm{k}_{\parallel} \cdot \bm{\rho})} \:,
\end{equation}
where $\bm{\rho}= (x,y)$ is a coordinate in the plane of the barrier. The function $u(z)$ in the regions I (incoming and reflected waves, see Fig.~1), II and III (transmitted wave)  has the form
\begin{equation}
u^{(I)}_{\pm}(\bm{r}) = \left[ \exp{(i k_z z)} + r_{\pm} \exp{(-i
k_z z)} \right] \:,
\end{equation}
\[
u^{(II)}_{\pm}(\bm{r}) = \left[ A_{\pm} \exp{( q_{\pm} z)} + B_{\pm} \exp{(-q_{\pm} z)} \right] \:,
\]
\[
u^{(III)}_{\pm}(\bm{r}) = t_{\pm} \exp{( i k_z z)} \:,
\]
respectively. Here $t_{\pm}$ and $r_{\pm}$ are the  transmission
and reflection coefficients for spin states $\chi_{\pm}$,
respectively, and the wave vectors under the barrier $q_{\pm}$ are
given by
\begin{equation}
q_{\pm}=q_0 \left( 1 \pm \gamma \frac{2 m_2 k_{\parallel}}{\hbar^2} \right)^{-1/2} \:,
\end{equation}
where $q_0$ is the reciprocal length of decay of the wave function
in the barrier for the case when the spin-orbit
interaction~(\ref{H_D}) is neglected
\begin{equation}
q_0 = \sqrt{\frac{2m_2 V}{\hbar^2} - k_z^2 \, \frac{m_2}{m_1} -
k_{\parallel}^2 \left( \frac{m_2}{m_1} -1 \right) } \:,
\end{equation}
and $m_i$ ($i=1,2$) are the effective masses outside and inside
the tunnelling barrier, respectively. Taking into account the
boundary conditions, which require that
\begin{equation}
u_{\pm} \hspace{1cm} \mbox{and}    \hspace{1cm} \frac{1}{m}\frac{
\partial u_{\pm}}{ \partial z}
\end{equation}
are continuous at  the interfaces, a system of linear equations
for $t_{\pm}$, $r_{\pm}$, $A_{\pm}$, and $B_{\pm}$ can be derived.
Solution of the system allows one to calculate the coefficients of
the transmission $t_{\pm}$. For the real case $ m_2 k_{\parallel}
\gamma / \hbar^2 \ll 1$ they are derived to be
\begin{equation}\label{tpm}
t_{\pm}=t_0 \exp{ \left( \pm \, \gamma  \frac{m_2 k_{\parallel}}{\hbar^2} \, aq_0  \right) } \:,
\end{equation}
where $t_0$ is the transmission coefficient when the spin-orbit
interaction~(\ref{H_D}) is neglected,
\begin{equation}\label{t0}
t_0=-4i \, \frac{m_2}{m_1} \frac{k_z q_0}{\left( q_0 - ik_z m_2/m_1 \right)^2} \exp{(-a q_0)} \:,
\end{equation}
$a$ is the width of the barrier. The general problem of tunnelling
of an electron with arbitrary initial spinor $\chi$ can be solved
by expanding $\chi$ to the eigen spinors $\chi_{\pm}$.

It is convenient to introduce a polarization efficiency ${\cal P}$
that determines the difference of tunnelling transmission
probabilities for the spin states "$+$" and "$-$" through the
barrier
\begin{equation}
{\cal P} = \frac{|t_+|^2 - |t_-|^2}{|t_+|^2 + |t_-|^2} \:.
\end{equation}
In our case it has the form
\begin{equation}\label{P}
{\cal P}= \tanh{ \left(  2 \gamma  \frac{m_2 k_{\parallel}}{\hbar^2} \, aq_0  \right) } \:.
\end{equation}

\begin{figure}[t]
\epsfxsize=3.7in \epsfysize=2.5in \centering{\epsfbox{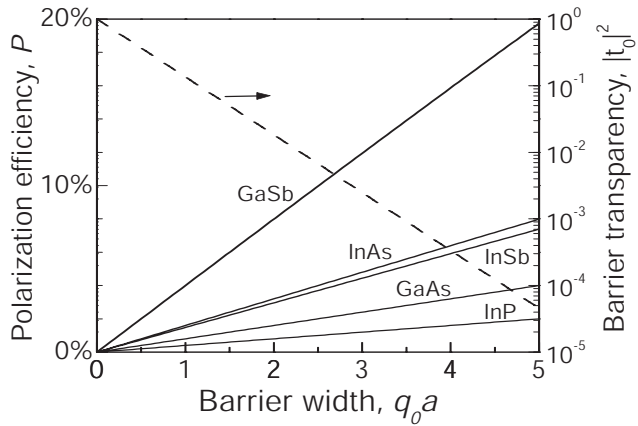}}
\caption{Coefficient of the polarization efficiency ${\cal P}$ as
a function of barrier width, $aq_0$, for various barrier
materials, and $k_{\parallel}=2 \cdot 10^{6}\,cm^{-1}$. }
\label{fig3}
\end{figure}

At a given initial wave vector of electrons, $\bm{k}$, the
polarization efficiency drastically increases with the strength of
the Dresselhaus spin-orbit coupling $\gamma$ (see Eq.~\ref{P}),
and the barrier width $a$. However while increase the barrier
width $a$ increases the tunnelling efficiency one should keep in
mind that the barrier transparency decreases simultaneously~(see
Eq.~(\ref{t0})). In Fig.~\ref{fig3} the efficiency ${\cal P}$ and
the barrier transparency $|t_0|^2$ are plotted as a function of
the barrier width, $q_0 a$, for various barrier materials.
The material parameters $\gamma$ and effective mass $m^* = m_2$
used in the calculations are given in the Table~1. One can see
that for a reasonable set of parameters given in the figure
caption it is possible to achieve spin polarizations of several
percents. Tunnelling barriers prepared  on the basis of GaSb or
its solutions seem to be the most efficient barrier materials for
spin selective tunnelling because of the large value of the
product $\gamma m^*$.

The polarization strongly depends on the electron wave vector
$k_\parallel$ parallel to the barrier (see Eq.~(\ref{P})). This
result suggests a device for spin injection into quantum wells.
Let's assume two quantum wells separated by a tunneling barrier,
and a current flowing along one of the quantum wells. The in-plane
current results in non-zero average electron wave vector
$k_{\parallel}$ and, due to the considered effect, in a spin
polarization of carriers.

In conclusion, we have demonstrated that the $k^3$ Dresselhaus
terms in the effective Hamiltonian of  semiconductors lacking a
center of inversion yield  a considerable spin polarization of
electrons tunnelling through barriers. The effect could be
employed for creating spin filters, eg. on the base of type-II
strained heterostructures like InAs/GaSb, InSb/GaSb and GaSb/GaAs.

The work was supported by the RFBR, the Presidium of the RAS,
and grants of the DFG and INTAS.


\end{document}